\documentclass[%
 reprint,
 amsmath,amssymb,
 aps,
]{revtex4-1}

\usepackage{graphicx}
\usepackage{subfigure}
\usepackage{dcolumn}
\usepackage{bm}

\begin{document}

\preprint{APS/123-QED}

\title{General bounds for quantum discord and discord distance}
\author{Feng Liu,$^{1,2}$~Guo-Jing Tian,$^{1}$~Su-Juan Qin,$^{1}$~Qiao-Yan Wen,$^{1}$~and Fei Gao$^{1}$}%
\email[ ]{gaofei$_$bupt@hotmail.com}
\affiliation{%
 $^{1}$State Key Laboratory of Networking and Switching Technology, Beijing University of Posts and Telecommunications, Beijing, 100876, China\\
 $^{2}$School of Mathematics and Statistics Science, Ludong University, Yantai, 264025, China}%

\begin{abstract}
For any bipartite state, how strongly can one subsystem be quantum correlated with another? Using the Koashi-Winter relation, we study the upper bound of purified quantum discord, which is given by the sum of the von Neumann entropy of the unmeasured subsystem and the entanglement of formation shared between the unmeasured subsystem with the environment. In particular, we find that the Luo et al.'s conjecture on the quantum correlations and the Lindblad conjecture are all ture, when the entanglement of formation vanishes. Let the difference between the left discord and the right discord be captured by the discord distance. If the Lindblad conjecture is true, we show that the joint entropy is a tight upper bound for the discord distance. Further, we obtain a necessary and sufficient condition for saturating upper bounds of purified quantum discord and discord distance separately with the equality conditions for the Araki-Lieb inequality and the Lindblad conjecture. Furthermore, we show that the subadditive relation holds for any bipartite quantum discord.
\begin{description}
\item[PACS numbers]
03.65.Ud, 03.65.Ta, 03.67.Mn
\end{description}
\end{abstract}

\pacs{Valid PACS appear here}
\maketitle


\section{Introduction}

Entanglement has always been identified as a key ingredient in quantum information processing, and it can be used to perform
certain tasks more efficiently than with classical correlation
only. However, entanglement does not account for all the nonclassical properties of quantum correlations. This is because 1) Knill and Laflamme [1] showed that quantum computation, in which a collection of qubits in a completely mixed state couple to a single control qubit that has nonzero purity, can achieve an exponential improvement in efficiency over classical computers for a limited set of tasks; 2) within the context of the quantum
measurement problem, Zurek and Vedral [2] concluded that even separable states usually contain correlations which are not entirely classical. These correlations are aptly named quantum discord ($QD$). As a measure of quantum correlations, $QD$ encapsulates entanglement but goes further, as it is also present even in separable states. Intriguingly, this could
be of practical significance because $QD$ is more easily produced and maintained than entanglement [3]. Over the past decade, $QD$ has been the focus of several theoretical and experimental studies addressing its formal characterization [4], its connection with entanglement distribution [5], remote state preparation [6] and unambiguous quantum state discrimination [7].

Despite the significance, the value of $QD$ is notoriously difficult to calculate due to the optimization procedure involved. Analytical results are known only for certain special
classes of states [8]. Particularly, it has been proved that
it is impossible to obtain a closed expression for $QD$, even for
general states of two qubits [9]. This fact makes it desirable to
obtain some computable bounds for $QD$, and several attempts
have been devoted to this issue in the past few years [10-12].

In 2010, Luo et al. [10] conjectured that both the classical and quantum correlations are (like the classical mutual information) bounded above by each subsystem's entropy. The conjecture seems intuitively reasonable since the marginal entropies quantify the effective sizes of the two subsystems in view of the Schumacher noiseless coding theorem [13], and some further supporting evidences for the conjecture are given [10]. On the conjecture of classical correlations in bipartite states, it has been rigorous proved that it is upper bounded by the von Neumann entropies of each subsystem in [14-16]. In any bipartite state, the quantum mutual information can be separated into two parts: classical correlation and $QD$, and it is considered as the total amount of correlations in the bipartite state. Under these conditions, it has been proved [14,16-17] that $QD$ is upper bounded by the von Neumann entropy of the measured subsystem. Recently, Xi et al. [11] revisited the upper bound of $QD$ given by the von Neumann entropy of the measured subsystem using a tradeoff between the amount of classical correlation and $QD$ in a tripartite pure state.

However, it has remained an open question as to what is the size relationship of $QD$ and the von Neumann entropy of the unmeasured subsystem in any bipartite state (or the corresponding tripartite pure state). For this question, some class of states have been given in Refs. [11,16], which conform Luo et al.'s conjecture [10]. (From now on, when we refer to the Luo et al.'s conjecture we mean the quantum correlation is bounded by each subsystem's entropy.) Recent investigations [14,18] give compelling evidence that there exist some bipartite states to disprove the conjecture. These facts motivate us to systematically investigate the upper bound of $QD$. And, it is interesting to give a necessary and sufficient condition for saturating the upper bound of $QD$. On the other hand, $QD$ is generally not symmetric, i.e. the left discord and the right discord are unequal, which may be expected because conditional entropy is not symmetric [19]. The difference between the left discord and the right discord is captured by the discord distance. Then, how close is the discord distance?

By purifying the bipartite quantum systems and using the Koashi-Winter relation [20], we prove that, for every bipartite systems, $QD$ cannot exceed the sum of the entropy of the unmeasured subsystem and entanglement of formation shared between the unmeasured subsystem with the environment. We further prove that the Luo et al.'s conjecture on the quantum correlations [10] and the Lindblad conjecture [21] are all true, when entanglement of formation vanishes. When the Lindblad conjecture is true, we show that the discord distance is upper bounded by the joint entropy. Intriguingly, we find that the subadditivity is a new important property of tripartite quantum discord.

This paper is organized as follows. In Sec. 2, we give a brief
review on $QD$ and Luo et al.'s conjecture [10]. In Sec. 3, we show that a subsystem being quantum correlated with another one limits its possible entanglement of formation with the environment, and then we prove Luo et al.'s conjecture when the entanglement of formation vanishes. In Sec. 4, we show that the Lindblad conjecture can be always true for a class of states, and then we prove the discord distance is upper bounded by the joint entropy. In Sec. 5, we show that for any tripartite state the subadditivity holds. Section 6 is the conclusion.
\section{Review of quantum discord}

Two systems $A$ and $B$ are correlated if together they contain more information than taken separately. If $A$ and $B$ are classical systems and we measure the lack of information by entropy, correlations between two random variables of them are in information theory quantified by the mutual information
\begin{equation}
\mathcal I(A:B)=\mathcal H(A)+\mathcal H(B)-\mathcal H(A,B)
\end{equation}
where $\mathcal H(\cdot)$ stands for the Shannon entropy. For quantum systems $A$ and $B$, i.e., Hilbert space $\mathcal H_{AB}=\mathcal H_{A}\otimes \mathcal H_{B}$ is a bipartite quantum composite system, and a bipartite quantum state $\rho^{AB}$ (density matrix) of the composite system, the total amount of correlations is quantified by quantum mutual information [22] between the two subsystems $A$ and $B$
\begin{equation}
\mathcal I(\rho^{AB})=S(\rho^{A})+S(\rho^{B})-S(\rho^{AB})
\end{equation}
where $S(\cdot)$ is the von Neumann entropy and $\rho^{A(B)}=Tr_{B(A)}(\rho^{AB})$ are reduced density matrices.

In contradistinction to the classical case, in the quantum
analog there are many different measurements that can be
performed on a subsystem, and measurements generally disturb
the quantum state [19]. A measurement on subsystem $A$ is described
by a positive-operator-valued measure (POVM) with
elements $E_{k}^{A}$, where $k$ is the classical outcome. In analogy with Eq. (2) and to quantify the classical correlations (on $B$) of the state $\rho^{AB}$ independently of a measurement on $A$, Vedral et al. [19] define an alternative version of the quantum mutual information
\begin{equation}
 \mathcal J_{A}(\rho^{AB})=S(\rho^{B})-min_{\{E_{k}^{A}\}}S(B|\{E_{k}^{A}\})
\end{equation}
where the minimum is taken over all POVM measurements $\{E_{k}^{A}\}$ on $A$, and $S(B|\{E_{k}^{A}\})=\sum_{k}p_{k}S(\rho^{B|k})$ is the averaged conditional von Neumann entropy of the nonselective postmeasurement state $\rho^{B|k}=Tr_{A}(E_{k}^{A}\rho^{AB})/p_{k}$ with $p_{k}=Tr(E_{k}^{A}\rho^{AB})$. Then, $QD$ (on $B$) of a state $\rho^{AB}$ under a measurement $\{E_{k}^{A}\}$ is defined as a difference between the total correlation, as given by the quantum mutual information in Eq. (2), and the classical correlation Eq. (3) [2,19]:

~~~~~~~~~~~~~$\mathcal D_{A}(\rho^{AB})=\mathcal I(\rho^{AB})- \mathcal J_{A}(\rho^{AB})$
\begin{equation}
=min_{\{E_{k}^{A}\}}S(B|\{E_{k}^{A}\})-S(B|A)
\end{equation}
where $S(B|A)=S(\rho^{AB})-S(\rho^{A})$ denotes the conditional von Neumann entropy of $\rho^{AB}$ [22], and the
minimum is taken over all POVM measurements $\{E_{k}^{A}\}$. $QD$ is not symmetric, i.e., in general $\mathcal D_{A}(\rho^{AB})\neq \mathcal D_{B}(\rho^{AB})$, which can be interpreted in terms of the probability of confusing certain quantum states [19]. Here, $\mathcal D_{A}(\rho^{AB})$ refers to the left discord, while $\mathcal D_{B}(\rho^{AB})$ refers to the right discord. From now on, when we refer to the discord we mean the left discord $\mathcal D_{A}(\rho^{AB})$.

There are interesting points as follows

~~~~~~~~~~~~~~$\mathcal I(A:B)\leq min\{\mathcal H(A),\mathcal H(B)\},$
\begin{equation}
 \mathcal I(\rho^{AB})\leq 2\cdot min\{S(\rho^{A}),S(\rho^{B})\}.
\end{equation}
In particular, if $\rho^{AB}=|\psi\rangle^{AB}\langle\psi|$ is a pure state, then $\mathcal I(\rho^{AB})=2S(\rho^{A})=2S(\rho^{B})$, which saturates Eq. (5). Now the total correlation $\mathcal I(\rho^{AB})$ is separated into classical correlation $\mathcal J_{A}(\rho^{AB})$ and quantum correlation $\mathcal D_{A}(\rho^{AB})$. Then, Eq. (5) can be recast into $\mathcal J_{A}(\rho^{AB})+\mathcal D_{A}(\rho^{AB})\leq 2\cdot min\{S(\rho^{A}),S(\rho^{B})\}$. From the perspective of correlative capacities, Luo et al. naturally proposed the following conjectures on the quantum correlation in a bipartite state $\rho^{AB}$, which splits the preceding inequalities [10,14]:

~~~~~~~~~~~~~~~~$\mathcal D_{A}(\rho^{AB})\leq S(\rho^{A})$~~~~~~~~~~~~~~~~~~~~~~~~~~~~~~~~~~~~~~~~(I)

~~~~~~~~~~~~~~~~$\mathcal D_{A}(\rho^{AB})\leq S(\rho^{B})$~~~~~~~~~~~~~~~~~~~~~~~~~~~~~~~~~~~~~~~~(II)

To illuminate the progress of the two conjectures, we first introduce two lemmas as follows.

\emph{Lemma 1} [16]. For any bipartite state $\rho^{AB}$, the $QD$ satisfies\\
(i) $\mathcal D_{A}(\rho^{AB})\leq S(\rho^{A})$;\\
(ii) $\mathcal D_{A}(\rho^{AB})\leq min\{S(\rho^{A}),S(\rho^{B})\}$, whenever $S(\rho^{A})\leq S(\rho^{B})$ or $\rho^{AB}$ is separable.

It is clear that Zhang and Wu [16] have proved Conjecture (I) after taking over all von Neumann measurements on the subsystem $A$. They also found that $\mathcal D_{A}(\rho^{AB})$ is upper bounded by the von Neumann entropy of the subsystem $B$ for a class of states. However, the constraints on the class of states in (ii) are so strong that all entangled pure states satisfying Conjecture (II) are excluded. It is necessary to consider relaxing the constraints in \emph{Lemma 1}. Using the method of
purification and a tradeoff relation between $QD$ and classical correlation, Xi et al. [11] got the following result.

\emph{Lemma 2} [11]. For any bipartite state $\rho^{AB}$, there is always a tripartite pure state $\rho^{ABC}=|\psi\rangle^{ABC}\langle\psi|$ such that $\rho^{AB}=Tr_{C}(\rho^{ABC})$, and the $QD$ satisfies\\
(i) $\mathcal D_{A}(\rho^{AB})\leq S(\rho^{A})$, with equality if and only if $S(\rho^{B})-S(\rho^{A})=S(\rho^{C})$;\\
(ii) $\mathcal D_{A}(\rho^{AB})=S(\rho^{B})$ if the equality $S(\rho^{A})-S(\rho^{B})=S(\rho^{C})$ is satisfied.

Taking over all POVM measurements on the subsystem $A$, Conjecture (I) is proved. And, the above sufficient condition (ii) for the situation of $\mathcal D_{A}(\rho^{AB})=S(\rho^{B})$ is also true for a class of states, including all entangled pure states as shown by \emph{Example 2} in Sec. 3.

The conclusions above appear to indicate that conjecture (II) is true. However, Li and Luo [14] pointed out that the
inequality (II) is not valid in general and provided a counter-example [18]:

\emph{Example 1}. Let $\rho^{AC}$ be a Werner state\\ $\rho^{AC}=\frac{d-x}{d^{3}-d}\mathcal I^{AC}+\frac{dx-1}{d^{3}-d}
\sum_{i,j=1}^{d}|ij\rangle^{AC}\langle ij|$\\
acting on $\mathcal C^{d}\otimes \mathcal C^{d}$, where $x\in (-1,0)$, $d\geq6$, $\{|ij\rangle \}$ is an orthogonal basis of product states for the composite system. If $\rho^{ABC}$ is a purification of $\rho^{AC}$, then $\mathcal D_{A}(\rho^{AB})>S(\rho^{B})$, which is quite counterintuitive.
\section{The upper bound of purified discord}
For any general bipartite state $\rho^{AB}$, we study bipartite $QD$ in the tripartite purified system $\rho^{ABC}$ such that $\rho^{AB}=Tr_{C}(\rho^{ABC})$. There is a Koashi-Winter relation [20]
\begin{equation}
\mathcal E_{F}(\rho^{BC})+\mathcal J_{A}(\rho^{AB})=S(\rho^{B}),
\end{equation}
which is a tradeoff between entanglement of formation and classical correlation. Here $\mathcal E_{F}(\rho^{BC})$ is entanglement of formation, defined as $\mathcal E_{F}(\rho^{BC})=min_{\{p_{i},|\psi_{i}\rangle\}}\Sigma_{i}p_{i}
S[Tr_{C}(|\psi_{i}\rangle\langle\psi_{i}|)]$, and the
minimum is taken over all pure ensembles $\{p_{i},|\psi_{i}\rangle\}$ satisfying $\rho^{BC}=\Sigma_{i}p_{i}|\psi_{i}\rangle\langle\psi_{i}|$. Using the interplay between mutual information and $QD$, Xi et al. [11] obtained the monogamic relation between $QD$ and the classical correlation as follows
\begin{equation}
\mathcal D_{A}(\rho^{AB})+\mathcal J_{A}(\rho^{AC})=S(\rho^{A})
\end{equation}
which is universal for any tripartite pure states $\rho^{ABC}$. The equation tells us that the amount of quantum correlation between $A$ and $B$, plus the amount of classical correlation between $A$ and the environment $C$, must be equal to the entropy of the measured subsystem $A$. In particular, the monogamic relation (7) directly supplies a general upper bound for $QD$, i.e., the Conjecture (I), which has been proved in Refs. [11,14,16]. Analogously, substituting $\mathcal J_{A}(\rho^{AB})=\mathcal I(\rho^{AB})-\mathcal D_{A}(\rho^{AB})$ and $-S(B|A)=\mathcal I(\rho^{AB})-S(B)$ into Eq. (6), we can also obtain the tradeoff [24] between $QD$ and the entanglement of formation as follows
\begin{equation}
\mathcal D_{A}(\rho^{AB})-\mathcal E_{F}(\rho^{BC})=-S(B|A)
\end{equation}
which gives an operational interpretation of the negative conditional entropy as the difference between the two non-classical correlations.

By applying the Araki-Lieb inequality to Eq. (8), we get a general upper bound for $QD$ as the following result, and determine which states saturate this bound.

\emph{Theorem 1}. For any bipartite state $\rho^{AB}$, $\rho^{ABC}$ is a purification of it, then we have
\begin{equation}
\mathcal D_{A}(\rho^{AB})\leq S(\rho^{B})+\mathcal E_{F}(\rho^{BC})
\end{equation}
with equality if and only if $S(\rho^{A})-S(\rho^{B})=S(\rho^{C})$.

This result shows that the necessary and sufficient conditions for saturating the upper bound of $QD$ in \emph{Theorem 1} and the result (ii) in \emph{Lemma 2} are consistent results. The causes of consistent results is attributed to $S(\rho^{A})-S(\rho^{B})= S(\rho^{C})$ and $\rho^{BC}=\rho^{B}\otimes \rho^{C}$ are equivalent [11].

As an illustration of the necessary and sufficient condition, let us consider the following two examples.

\emph{Example 2}. Let $\rho^{AB}$ is a pure state. For any bipartite pure state, $S(\rho^{A})=S(\rho^{B})$ and $QD$ is just the entropy of the reduced states of the subsystems. Using the fact that a entropy is zero if and only if the state is pure, we get the equality condition of the Araki-Lieb inequality $S(\rho^{A})-S(\rho^{B})= S(\rho^{AB})=0$. On the base of the \emph{Theorem 1} and \emph{Lemma 2}, $\mathcal D_{A}(\rho^{AB})=\mathcal D_{B}(\rho^{AB})=S(\rho^{B})$ hold.

If $\rho^{ABC}$ is a purification of $\rho^{AB}$, we now apply Eq. (5) to the tripartite pure state $\rho^{ABC}=|\psi\rangle^{ABC}\langle \psi|$, and get $\mathcal I(\rho^{ABC})\leq 2\cdot S(\rho^{AB})=0$. Then, $\rho^{ABC}=\rho^{AB}\otimes\rho^{C}$. So we have $\mathcal J_{B}(\rho^{BC})=0$, and  obviously $\mathcal E_{F}(\rho^{BC})=0$. These results show that the condition of $\mathcal D_{A}(\rho^{AB})=S(\rho^{A})$ follows from \emph{Theorem 1} in Ref. [11], and the condition of $\mathcal D_{A}(\rho^{AB})=S(\rho^{B})+E_{F}(\rho^{BC})$ follows from our \emph{Theorem 1}.

\emph{Example 3}. As given in Ref. [11], let

$\rho^{AB}=\frac{1}{4}[
(|00\rangle^{A}\langle00|+|01\rangle^{A}\langle01|)\otimes|0\rangle^{B}\langle0|+(|00\rangle^{A}\langle10|+|01\rangle^{A}\langle11|)\otimes |0\rangle^{B}\langle1|+$

~~~~~~~~~~$(|10\rangle^{A}\langle00|+|11\rangle^{A}\langle01|)\otimes |1\rangle^{B}\langle0|+
(|10\rangle^{A}\langle10|+|11\rangle^{A}\langle11|)\otimes |1\rangle^{B}\langle1|]$\\
where $\mathcal H^{A}=\mathcal C^{2}\otimes \mathcal C^{2}$, $\mathcal H^{B}=\mathcal C^{2}$. Notice that this state is a mixed state, since $Tr[(\rho^{AB})^{2}]=\frac{1}{2}<1$. We can always
find a tripartite pure state $\rho^{ABC}=|\Psi\rangle^{ABC}\langle\Psi|$ such that $\rho^{AB}=Tr_{C}(\rho^{ABC})$ and $|\Psi\rangle^{ABC}=\frac{1}{\sqrt{2}}(|\psi_{0}\rangle^{AB}|0\rangle^{C}+|\psi_{1}\rangle^{AB}|1\rangle^{C})$, where $|\psi_{0}\rangle=\frac{1}{\sqrt{2}}(|00\rangle^{A}|0\rangle^{B}+|10\rangle^{A}|1\rangle^{B})$ and
$|\psi_{1}\rangle=\frac{1}{\sqrt{2}}(|01\rangle^{A}|0\rangle^{B}+|11\rangle^{A}|1\rangle^{B})$.
The reduced states can be obtained $\rho^{A}=\frac{I^{A}}{4}$ and $\rho^{B}=\frac{I^{B}}{2}$, where $I^{A}$ and $I^{B}$ are respectively identity operators on $\mathcal H^{A}$ and $\mathcal H^{B}$. After some calculations one obtains $S(\rho^{A})=2$,~$S(\rho^{AB})=S(\rho^{B})=1$.

Using the equality $\mathcal D_{A}(\rho^{AB})=\mathcal E_{F}(\rho^{BC})+S(\rho^{A})-S(\rho^{AB})$, we know the upper bound in Eq. (9) is achievable if and only if $S(\rho^{A})-S(\rho^{AB})= S(\rho^{B})=1$.

From the above results, we know \emph{Theorem 1} with equality if and only if the equality in the Araki-Lieb inequality holds. And, the unmeasured subsystem cannot correlate with the
environment if $QD$ between $A$ and $B$
is equal to the entropy of the unmeasured subsystem. What is astonishing is that the result and \emph{Theorem 2} in Ref. [11]  are almost entirely the same with equality. However, the upper bound of $QD$ is obtained in a quite different form from Conjecture (II). To this end, we have the following sufficient condition for the situation of Conjecture (II), i.e., $\mathcal D_{A}(\rho^{AB})\leq S(\rho^{B})$.

\emph{Corollary 1}. For any bipartite state $\rho^{AB}$, $\rho^{ABC}$ is a purification of it and $\mathcal E_{F}(\rho^{BC})=0$, then we have
\begin{equation}
\mathcal D_{A}(\rho^{AB})\leq S(\rho^{B})
\end{equation}
with equality if and only if $S(\rho^{A})-S(\rho^{B})= S(\rho^{C})$.

The result shows that Conjecture (II) is true if the equality $\mathcal E_{F}(\rho^{BC})=0$ is satisfied. In other words, $QD$ between $A$ and $B$ is upper bounded by the entropy of the unmeasured subsystem if the unmeasured subsystem $B$ cannot entangled with the environment $C$. Further, the maximal $QD$ between $A$ and $B$ will even forbid $B$ from being correlated to other systems outside this composite system. One thing to be noted is $\mathcal {D}_A (\rho^{AB})=-S(B|A)$ by the Koashi-Winter equality when $\mathcal {E}_F(\rho^{BC})=0$, so we are just trying to formally establish the relation between discord and the unmeasured subsystem.
\section{The upper bound of discord distance}

The discord distance of a quantum state $\rho^{AB}$ under POVM measurements $\{E_{k}^{A}\}$ and $\{E_{k}^{B}\}$ is defined as a difference between the two one-side discords (left discord and right disocrd):
\begin{equation}
|\mathcal D_{A}(\rho^{AB})-\mathcal D_{B}(\rho^{AB})|.
\end{equation}
To be clearer, we substitute $\mathcal D_{i}(\rho^{AB})=\mathcal I(\rho^{AB})-\mathcal J_{i}(\rho^{AB})$, where $i=A, B$, and obtain the classical correlation distance $|\mathcal J_{A}(\rho^{AB})-\mathcal J_{B}(\rho^{AB})|$. A suitable physical interpretation of the discord distance is how close are two classical correlations for $A$ versus $B$ in the quantum state. The Eq. (11) shows that smaller difference implies more fair communication in an ideal quantum network. From \emph{Lemma 2} in Sec. 2 and \emph{Corollary 1} in Sec. 3, we have a sufficient condition for the
situation of $\mathcal D_{A}(\rho^{AB})=\mathcal D_{B}(\rho^{AB})=S(\rho^{B})$ if the equality $S(\rho^{A})-S(\rho^{B})=S(\rho^{AB})$ is satisfied. Meantime, $\mathcal J_{A}(\rho^{AB})=\mathcal J_{B}(\rho^{AB})=-S(B|A)$  can also be deduced from the above sufficient condition. Though, in its present form, a necessary and sufficient condition cannot be obtained. We leave it as an open question whether one could consider variations of the proof procedure that would render it
suitable for fair quantum communication tasks.

Combining the monogamic relation in Eq. (6) with the equality condition for the strong subadditivity inequality
\begin{equation}
S(B|A)+S(B|C)=0,
\end{equation}
we have the following result.

\emph{Theorem 2}. For any bipartite state $\rho^{AB}$, $\rho^{ABC}$ is a purification of it and $\mathcal E_{F}(\rho^{AC})=\mathcal E_{F}(\rho^{BC})=0$, then we have
\begin{equation}
|\mathcal D_{A}(\rho^{AB})-\mathcal D_{B}(\rho^{AB})|\leq S(\rho^{AB}).
\end{equation}

To prove this theorem, we first introduce the Lindblad conjecture [21], which states that the classical correlation account for at least half of the total correlation, or equivalently,
correlations are more classical than quantum.

\emph{Lindblad conjecture}. For any bipartite state $\rho^{AB}$, Lindblad proposed the following conjecture $\mathcal D_{A}(\rho^{AB})\leq \mathcal J_{A}(\rho^{AB})$, which is based on several intuitive observations [10,21].

Luo and Zhang [21] disproved the intuitive conjecture of Lindblad by evaluating an observable correlations for generic two-qubit states and obtain analytical expressions in some particular cases. And, they provided a counter-example:

\emph{Example 4}. Let $\rho^{AB}$ be a Werner state $\rho^{AB}=\frac{I^{AB}}{6}+\frac{1}{3}|\psi^{-}\rangle^{AB}\langle \psi^{-}|$, where $|\psi^{-}\rangle=\frac{1}{\sqrt{2}}(|01\rangle-|10\rangle)$.  In such a situation,

$\mathcal D_{A}(\rho^{AB})> \mathcal J_{A}(\rho^{AB})$, where $\mathcal D_{A}(\rho^{AB})\simeq 0.126$, $\mathcal J_{A}(\rho^{AB})\simeq 0.082$.

From \emph{Corollary 1}, we find that \emph{Lindblad conjecture} can be true, and introduce a lemma as follows.

\emph{Lemma 3}. For any bipartite state $\rho^{AB}$, $\rho^{ABC}$ is a purification of it and $\mathcal E_{F}(\rho^{BC})=0$, then we have $\mathcal D_{A}(\rho^{AB})\leq \mathcal J_{A}(\rho^{AB})$.

\emph{Proof}. From the \emph{Corollary 1} in Sec. 3, we get $\mathcal D_{A}(\rho^{AB})\leq S(\rho^{B})$ when $\mathcal E_{F}(\rho^{BC})=0$. While under the same condition, we have $\mathcal J_{A}(\rho^{AB})=S(\rho^{B})$ due to the Koashi-Winter relation $\mathcal J_{A}(\rho^{AB})+\mathcal E_{F}(\rho^{BC})=S(\rho^{B})$, i.e. Eq. (6). It turns out that \emph{Lemma 3} is true.

The result shows that $QD$ is always bounded from above by the classical correlation when the entanglement of formation, between the unmeasured subsystem $B$ and the environment $C$, vanishes. Further, the maximal $QD$ between $A$ and $B$ will forbid system $B$ from being correlated to other systems outside this composite system when \emph{Lindblad conjecture} is true. The phenomenon is  very real significance in quantum information theory. The accessible information is a measure of how well the receiver can do at inferring the information being included in the other subsystem [22]. And, the difference between the accessible information achieves the maximum of the classical correlation $\mathcal J_{B}(\rho^{BC})$ if and only if \emph{Lindblad conjecture} is true and $\mathcal E_{F}(\rho^{BC})=0$.

Using \emph{Lemma 3}, we will complete the proof of \emph{Theorem 2}.

\emph{Proof of Theorem 2}. Consider that $\mathcal D_{B}(\rho^{BC})\leq \mathcal J_{B}(\rho^{BC})$, i.e., $2\cdot \mathcal J_{B}(\rho^{BC})\geq \mathcal I(\rho^{BC})$, the inequality follows from the result of \emph{Lemma 3} with $\mathcal E_{F}(\rho^{AC})=0$. We have
\begin{equation*}
2\mathcal J_{B}(\rho^{BC})+S(B|C)\geq \mathcal I(\rho^{BC})+S(B|C)
\end{equation*}
\begin{equation}
=S(\rho^{B})\geq \mathcal J_{A}(\rho^{AB}).
\end{equation}
The first inequality follows by adding the item $S(B|C)$ to the left and right hand sides of $2\cdot \mathcal J_{B}(\rho^{BC})\geq \mathcal I(\rho^{BC})$. The first equality follows because the quantum mutual information has an equivalent form for the classical mutual information $\mathcal I(\rho^{BC})=S(\rho^{B})-S(B|C)$. The second inequality follows from the Eq. (6), which is the Koashi-Winter relation.

Using Eqs. (12) and (14), the classical correlation $\mathcal J_{A}(\rho^{AB})$ admits the following representation:
\begin{equation*}
\mathcal J_{A}(\rho^{AB})\leq 2\cdot\mathcal J_{B}(\rho^{BC})-S(B|A)
\end{equation*}
\begin{equation}
=2\cdot\mathcal J_{B}(\rho^{BC})+\mathcal I(\rho^{AB})-S(\rho^{B}).
\end{equation}
In terms of $QD$ and the von Neumann entropy of the corresponding unmeasured subsystem, Eq. (15) can be rephrased as follows.
\begin{equation}
S(\rho^{B})\leq 2\mathcal J_{B}(\rho^{BC})+\mathcal D_{A}(\rho^{AB}).
\end{equation}

In addition, we have
\begin{equation}
\mathcal D_{A}(\rho^{AB})\leq S(\rho^{B})
\end{equation}
from the \emph{Corollary 1} and $\mathcal E_{F}(\rho^{BC})=0$.

Using Eq. (7) and Eqs. (16-17), we have

$|\mathcal D_{A}(\rho^{AB})-\mathcal D_{B}(\rho^{AB})|=|\mathcal D_{A}(\rho^{AB})+\mathcal J_{B}(\rho^{BC})-S(\rho^{B})|\leq \mathcal J_{B}(\rho^{BC}).
$

Under the assumption $\mathcal E_{F}(\rho^{AC})=\mathcal E_{F}(\rho^{BC})=0$, it can be shown that $J_{A}(\rho^{AC})=J_{B}(\rho^{BC})=S(\rho^{AB})$ by using the Koashi-Winter relation for other permutations of $A$, $B$,
and $C$.

This completes the proof of \emph{Theorem 2}.

Note: In fact, this is a direct result of the Koashi-Winter equality (8) and the Araki-Lied inequality, see the following:

   $\mathcal {D}_A(\rho^{AB}) = \mathcal {E}_F(\rho^{BC}) - S(B|A),$
   $\mathcal {D}_B(\rho^{AB}) = \mathcal {E}_F(\rho^{AC}) - S(A|B).$\\
From the above two Koashi-Winter equalities one can obtain

   $\mathcal {D}_A(\rho^{AB}) -\mathcal {D}_B(\rho^{AB}) = S(\rho^A) - S(\rho^B),$\\
Using $\mathcal {E}_F(\rho^{AC}) = \mathcal {E}_F(\rho^{BC})=0$, we have $|\mathcal D_{A}(\rho^{AB})-\mathcal D_{B}(\rho^{AB})|\leq \mathcal J_{A}(\rho^{AB}).$ This completes the proof of \emph{Theorem 2}. Using the above detailed proof process, it is mainly because we want to show the Lindblad conjecture's value.

Further, we give the explicit characterization of the quantum
states saturating the upper bound of discord distance as follows.

\emph{Corollary 2}. For any bipartite state $\rho^{AB}$, $\rho^{ABC}$ is a purification of it and $\mathcal E_{F}(\rho^{AC})=\mathcal E_{F}(\rho^{BC})=0$, then we have
\begin{equation}
\mathcal D_{B}(\rho^{AB})-\mathcal D_{A}(\rho^{AB})\leq  S(\rho^{AB})
\end{equation}
with equality if and only if $\mathcal E_{F}(\rho^{BC})=0$ and $\mathcal D_{B}(\rho^{AB})=J_{B}(\rho^{AB})$.

\emph{Proof}. Eq. (18) is true which can easily be deduced from \emph{Theorem 2}.

Based on \emph{Lemma 3} and the Koashi-Winter relation, we can give a necessary and sufficient condition for the
situation of $\mathcal D_{B}(\rho^{AB})-\mathcal D_{A}(\rho^{AB})= S(\rho^{AB})$. The equality can be rephrased as follows.

$0=\mathcal D_{B}(\rho^{AB})-\mathcal D_{A}(\rho^{AB})-S(\rho^{AB})=S(\rho^{B})-2\mathcal J_{B}(\rho^{BC})-\mathcal D_{A}(\rho^{AB})$

~~$=S(\rho^{B})-2\mathcal J_{B}(\rho^{BC})-(\mathcal I(\rho^{AB})-\mathcal J_{A}(\rho^{AB}))$

~~$=\mathcal J_{A}(\rho^{AB})-S(B|C)-2\mathcal J_{B}(\rho^{BC})$.\\
For the second equality, by the Koashi-Winter relation $\mathcal D_{B}(\rho^{AB})+\mathcal J_{B}(\rho^{BC})=S(\rho^{B})$; for the second equality, by $\mathcal D_{A}(\rho^{AB})+\mathcal J_{A}(\rho^{AB})= \mathcal I(\rho^{AB})$; for the last equality, by $\mathcal I(\rho^{AB})=S(\rho^{B})-S(B|A)$ and Eq. (12).

From the result in Eq. (15), we know that $\mathcal J_{A}(\rho^{AB})=S(B|C)+2\mathcal J_{B}(\rho^{BC})$ if and only if $\mathcal E_{F}(\rho^{BC})=0$ and $\mathcal D_{B}(\rho^{AB})=J_{B}(\rho^{AB})$. This completes the proof of \emph{Corollary 2}.
\section{Subadditivity of discord}

Quantum discord has the following properties [19]: asymmetric, nonnegative, invariant under local-unitary transformations, and vanishes if and only if the related state is classical quantum. Furthermore, discord is bounded from above by the von Neumann entropy of the measured subsystem [11,12], and the sum of the von Neumann entropy of the unmeasured subsystem and the entanglement of formation shared between the unmeasured subsystem with the environment (\emph{Theorem 1} in Sec. 3). In 2012, Prabhu et al. [25] investigated the monogamy relationship for quantum discord. They showed that for any tripartite state $\rho^{ABC}$, the inequality
$\mathcal D_{B}(\rho^{AB})+\mathcal D_{C}(\rho^{AC})\leq \mathcal D_{BC}(\rho^{ABC})$ holds if and only if $I(A:B:C)\geq J_{BC}(\rho^{ABC})$, where $I(A:B:C)=I(A:B)-I(A:B|C)$.

Using the definition of discord, we prove a very important property relating discord to the subadditivity as follows.

\emph{Theorem 3}. For any tripartite state $\rho^{ABC}$, we have
\begin{equation}
\mathcal D_{BC}(\rho^{ABC})\leq \mathcal D_{B}(\rho^{ABC})+\mathcal D_{C}(\rho^{ABC}).
\end{equation}

\emph{Proof}. Let $\Pi_{B}^{*}$ and $\Pi_{C}^{*}$ be the optimal complete projective measurements over $B$ and $C$ for the sake of $\mathcal D_{B}(\rho^{ABC})$ and $\mathcal D_{C}(\Pi_{B}^{*}(\rho^{ABC}))$ separately. Then the inequality (19) is obtained as follows:

$\mathcal D_{BC}(\rho^{ABC})\leq S(\rho^{ABC}||\Pi_{B}^{*}\otimes \Pi_{C}^{*}(\rho^{ABC}))$

$=S(\Pi_{B}^{*}\otimes \Pi_{C}^{*}(\rho^{ABC}))-S(\rho^{ABC})$

$=[S(\Pi_{B}^{*}(\rho^{ABC}))-S(\rho^{ABC})]+[S(\Pi_{B}^{*}\otimes \Pi_{C}^{*}(\rho^{ABC}))-S(\Pi_{B}^{*}(\rho^{ABC}))]$

$=\mathcal D_{B}(\rho^{ABC})+S[ \Pi_{C}^{*}(\Pi_{B}^{*}(\rho^{ABC}))]-S(\Pi_{B}^{*}(\rho^{ABC}))]$

$=\mathcal D_{B}(\rho^{ABC})+\mathcal D_{C}(\Pi_{B}^{*}(\rho^{ABC}))\leq \mathcal D_{B}(\rho^{ABC})+\mathcal D_{C}(\rho^{ABC})$.

This completes the proof of \emph{Theorem 3}.

We now apply Eq. (19) to a tripartite pure state $\rho^{ABC}$. For any bipartite pure state, the relative entropy of entanglement and the relative
entropy of discord coincide with the entropy of the reduced
states of the parts [23]. Thus, Eq. (19) becomes

$S(\rho^{BC})\leq S(\rho^{B})+S(\rho^{C})$\\
which is the subadditivity [22] of entropy for subsystems $BC$. Accordingly, Eq. (19) can also be seen as a generalization of the
subadditivity of entropy valid for tripartite mixed states.
\section{Conclusion}

In this work, we prove that the Luo et al.'s conjecture [10] and
the Lindblad conjecture [21] are all true, when the entanglement of formation between the unmeasured subsystem and the
environment vanishes. Further, the maximal quantum discord will even forbid the unmeasured subsystem from being correlated
to the environment. We have also shown that the discord distance is always bounded from above by the amount of the joint entropy when the Lindblad conjecture is true. Though, there in no exemplary state which can show the meaningfulness of \emph{Theorem 2}, we leave it as an open question whether one could give an operational interpretation of it. Finally, we find that the subadditivity is a new important property of tripartite quantum discord. \\\\
\textbf{Acknowledgments} We thank Professor Shunlong Luo and Nan Li for their helpful discussions. This work is supported by NSFC (Grant Nos. 61300181, 61272057, 61202434, 61170270, 61100203, 61121061), Beijing Natural Science Foundation (Grant No. 4122054), Beijing Higher Education Young Elite Teacher Project (Grant Nos. YETP0475, YETP0477), and BUPT Excellent Ph.D. Students Foundation (Grant No. CX201434).

\bibliography{mybibfile}

\end{document}